\documentclass[12pt]{article}
\usepackage{amsmath,amssymb}
\usepackage{graphicx}
\usepackage{caption}
\usepackage{color}
\usepackage{dcolumn}
\usepackage{bm}
\usepackage[numbers,super,comma,sort&compress]{natbib}

\definecolor{background-color}{gray}{0.98}

\title{Quantum Confinement in Hydrogen Bond}
\author{C. S. dos Santos\thanks{Instituto de Bioci\^encias, Letras e Ci\^encias
Exatas, IBILCE-UNESP,  S\~ao Jos\'e do Rio Preto, SP, Brazil}, E. Drigo Filho\thanks{Instituto de Bioci\^encias, Letras e Ci\^encias
Exatas, IBILCE-UNESP,  S\~ao Jos\'e do Rio Preto, SP, Brazil}, R. M. Ricotta \thanks{Faculdade de Tecnologia de S\~ao Paulo, FATEC/SP-CEETPS-UNESP, S\~ao Paulo, SP, Brazil}}

\begin{document}

\maketitle

\def\br{\begin{eqnarray}}
\def\er{\end{eqnarray}}
\def\be{\begin{equation}}
\def\ee{\end{equation}}
\def\({\left(}
\def\){\right)}
\def\d{\delta}
\def\h{ {1\over 2}  }
\def\m{\mu}
\def\o{\over}

\begin{abstract}  In this work, the quantum confinement effect is proposed as the cause of the displacement of the vibrational spectrum of molecular groups that involve hydrogen bonds. In this approach the hydrogen bond imposes a space barrier to hydrogen and constrains its oscillatory motion. We studied the vibrational transitions through the Morse potential, for the NH and OH molecular groups inside macromolecules in situation of confinement (when hydrogen bonding is formed) and non-confinement (when there is no hydrogen bonding). The energies were obtained through the variational method with the trial wave functions obtained from  Supersymmetric Quantum Mechanics (SQM) formalism. The results indicate that it is possible to distinguish the emission peaks related to the existence of the hydrogen bonds. These analytical results were satisfactorily compared with experimental results obtained from infrared spectroscopy.
\end{abstract}

\clearpage

\clearpage
  \makeatletter
  \renewcommand\@biblabel[1]{#1.}
  \makeatother

\bibliographystyle{apsrev}

\renewcommand{\baselinestretch}{1.5}
\normalsize

\clearpage

\section*{\sffamily \large INTRODUCTION} 
Hydrogen bond is an important interaction present in different systems in nature; the most famous examples of its occurrence are in water molecules and the DNA double helix, \cite{Arunan1} and references therein. Particularly, biological macromolecules as proteins, nucleic acids and carboxylic acids have the molecular groups NH and OH in their structures. These groups induce the hydrogen bonds which are essential to stabilize the structures and play an important role in the behavior of these macromolecules. Several other systems like clathrates and liquid crystals present hydrogen bonds in their structures \cite{Jeffrey}. 

Few years ago a modern definition of the hydrogen bond following a list of criteria was recommended, after a survey on recent results of the subject, \cite{Arunan2}.  According to this definition the hydrogen bond is an attractive interaction between a hydrogen atom from a molecule or a molecular fragment X-H in which X is more electronegative than H, and an atom or a group of atoms in the same or a different molecule, in which there is evidence of bond formation. The hydrogen bond can be classified as a donor-acceptor interaction.  The notation of the bond is X-H$\cdots$Y, where the dots represent the bond,  X-H is the hydrogen bond donor and Y is the acceptor anion. The covalent bond is represented by X-H and the hydrogen bond is represented by H$\cdots$Y. The high electro-negativity of atom X compared with the hydrogen atom in the X-H chemical bond makes the hydrogen to share part of its electronic density. 

The different spectroscopy techniques, such as, Raman, infrared (IR), Terahertz (THz) and Nuclear Magnetic Resonance  (NMR) spectroscopy,  provide the identification of the vibrational transitions of the molecular groups involved, since they are sensitive to changes in the spectrum. Therefore, it is possible to know if a specific molecular group participate of the hydrogen bond, because the vibrational spectrum changes. 

In this work it is proposed that quantum confinement effect is the main cause of de\-localization of the vibrational spectrum of NH and OH groups when hydrogen bonds are formed. 

Quantum confinement can be defined as the restricted space where the particles can move and it is generally introduced by a potential barrier  created, for instance, due to electrostatic potentials, the presence of an interface between different  materials or the presence of a surface. The simplest case of confinement consists of a particle subject to an infinite square well potential treated in introductory quantum mechanics, \cite{Griffiths}. In this confined regime, the motion of a particle is restricted to a well defined region of space and there is the appearance of discrete energy levels due to quantization effects. They appear once the size of the barrier is of the same magnitude as the de Broglie wavelength of the particle wave function. Confined systems have interesting applications; in material science the quantum confinement effect  is well studied theoretically and experimentally in semiconductor physics, mainly in the study of quantum dots, \cite{Ren}, \cite{Drigo1}. If the size of the material is reduced, the electronic and optical properties are drastically changed and can be controlled; this is one of the main subjects of nanostructure science, \cite{Harrison}. There are other cases that may also be studied by confinement as, for instance, a system subjected to high pressure, \cite{Kangarlu}.  Recent studies focus on theoretical as well as experimental investigations of hydrogen/halogen bonds in different confining environments, \cite{Lipkowski} - \cite{Ajami}.

In the approach used in this work the confinement is introduced by a potential barrier. In the formation of the hydrogen bond, the acceptor atom of the hydrogen bond introduces a barrier which prevents the free oscillation of the hydrogen atom attached to the donor atom. In this situation, the hydrogen is confined between two atoms and the energy levels are distorted by this confinement, i. e., there are vibrational frequency shifts. Vibrational frequency shifts can also be induced by high pressure, \cite{Wiederkehr} -\cite{Zakin2}.  

Thus, the main interest of this paper is to study the hydrogen bonds present in biological macromolecules, particularly, to determine the vibrational transitions related to OH and NH groups. To overcome this proposal we used an analytical methodology based on the variational method associated to Supersymmetric Quantum Mechanics, (SQM).

The confined quantum mechanical problem is settled  by considering the Schr$\ddot {o}$dinger equation depending on a potential with a potential barrier. Considering that the hydrogen bonds are highly directional, the vibrations of the groups studied are better described by the one-dimensional Morse potential.  If not by the potential barrier (without confinement) this potential presents exact analytical solution, \cite{Drigo2}. However, by considering the confined case (with confinement) the problem is not exactly solvable and therefore an approximation method has to be introduced.  At this point we use the formalism of SQM to calculate the spectrum of wave functions and energies and use the variational method to treat the confined case, with a potential barrier.  The transition energy between two adjacent states is calculated and the values obtained are compared with experimental spectroscopy data to validate the proposed hypothesis. The formalism of SQM coupled to the variational method has been a successful methodology to study many problems, in particular, the confined systems, \cite{Drigo1}, \cite{Silva}-\cite{Varshni}.

Other theoretical models based on the use of the Morse potential have been proposed to describe high frequency and low frequency vibration modes of the hydrogen bonds, \cite{Leviel}-\cite{Blaise}. In the approach proposed here, unlike these models, there is no need either to use any coupling theory nor to enter new parameters in the Morse potential. The data used here refer only to the covalent bonding and to the length of the hydrogen bond, which reduces significantly the number of parameters. 

\section*{\sffamily \large MODEL}
A schematic model of a hydrogen bond between atoms X and Y is shown in Figure 1.

\noindent In a hydrogen bonding, the hydrogen nuclei may oscillate within a region determined between a minimum distance ($x_{min}$) and a  maximum distance ($x_{max}$). The value of $x_{min}$ is given by the covalent radius ($x_c$) of atom X and $x_{max}$ is given by the length of the hydrogen bond ($l_{x \cdots y}$) subtracted by the van der Waals radius ($x_{vw}$) of the acceptor atom, i.e., 
\be
\label{xmax}
x_{max} = l_{X \cdots Y} - x_{vw}
\ee
The origin of the coordinate system is located at the center of the atom X. Any distance used is computed from this point and it is considered center-to-center. The covalent radius is adopted by the condition of impenetrability of the hydrogen atom donor; the employment of the van der Waals radius is justified because it is a measure of the volume excluded by the acceptor atom, i.e., the hydrogen cannot penetrate this region. The atoms are considered as hard spheres and the hydrogen is treated as a material point, since the electron cloud is shared with the donor and the acceptor atoms.

As mentioned before, the vibrational energy of the system is described through the confined one-dimensional Morse potential. This potential has been widely used to describe oscillations of diatomic molecules, \cite{Drigo2}-\cite{Ley-Koo}. The energy absorbed/emitted is obtained from the calculation of the energies of the ground state ($n = 0$) (for NH and OH), the first excited state ($n = 1$) (for NH and OH) and the second excited state ($n = 2$) (only for OH). The energy of the absorbed/emitted photon is obtained from the difference between two adjacent levels. The energies of the confined case are evaluated by using the variational method associated to the formalism of SQM.
\section*{\sffamily \large SUPERSYMMETRIC QUANTUM MECHANICS, SQM}
SQM is an algebraic method commonly used to exploit different aspects of non-relativistic quantum mechanical systems. It is particularly efficient to solve exactly all the quantum potential problems by means of the Hamiltonian factorization, thus providing the entire spectrum of wave functions with their respective energies together with  a hierarchy of Hamiltonians, all related by the supersymmetric algebra, \cite{Sukumar}. On the other hand, when the potential is non-exactly solvable, such as in a confined system, an approximation technique is needed and the variational method has already appeared as fully appropriate, \cite{Drigo2}.

Thus, using the superalgebra a given Hamiltonian can be factorized in terms of bosonic operators. In $\hbar = c = 1$ units, the   Schr$\ddot {o}$dinger equation can be rewritten as
\be
H_1 =  -{d^2 \o d x^2} + V_1(x) =  A_1^+A_1^-  + E_0^{(1)} 
\ee
where $ E_0^{(1)}$ is the lowest eigenvalue with ground state wave function satisfying 
\be
A_1^- \Psi_0^{(1)}(x)=0
\ee
and with the bosonic operators defined as
\be 
A_1^{\pm} =  \left(\mp {d \o dx} + w_1(x) \right) 
\ee
where the superpotential $w_1(x)$ satisfies the Riccati equation
\be
\label{Riccati}
w_1^2 - {d \o dx}w_1=  V_1(x) - E_0^{(1)}  .    
\ee
and the ground state wave function has the following form
\be
\label{eigenfunction}
\Psi_0^{(1)} (x) = N exp( -\int_0^x w_1(\bar x) d\bar x).
\ee

Thus a whole hierarchy of Hamiltonians can be constructed, with simple
relations connecting the eigenvalues and eigenfunctions of the $n$-members, 
\be
H_n = A_n^+A_n^- + E_0^{(n)} \;\;,
\ee
\be 
A_n^{\pm} =  \left(\mp {d \o dx} + w_n(x) \right) 
\ee
\be
\label{Psis}
\Psi_n^{(1)} = A_1^+A_2^+...\psi_0^{(n+1)}\;\;\;\;\;E_n^{(1)} = E_0^{(n+1)}.
\ee
In equation (\ref{Psis}), the upper index between parentheses refers to the hierarchy Hamiltonian member and the lower index to the level within the hierarchy. 
\section*{\sffamily \large MATHEMATICAL FORMALISM}
The one-dimensional Morse potential\cite{Drigo2}, is given by
\be
V_M = D_e (e^{-2\beta(x - x_{eq})} - 2e^{-\beta(x - x_{eq})})
\ee
where  $D_e $  is the dissociation energy of molecule, $\beta$  is the parameter related to the width of potential well and $x_{eq}$ is the inter-nuclear equilibrium distance. The $\beta$ parameter may be determined by treating the molecule as a frequency oscillator $\nu$ and through  experimental data. By expanding the Morse potential in a Taylor series around the equilibrium position to the second order we find that
\be
\label{beta}
\beta= ({2\pi^2 \mu\over D_e})^\h\nu\;.
\ee
The Schr$\ddot {o}$dinger equation for a molecule subject to the Morse potential, according to the model, is written in one-dimensional Cartesian coordinates with the  change of variables, $y = \beta x$. After multiplying both sides by  ${2 \mu\over \beta^2\hbar^2}$, we obtain the Schr$\ddot {o}$dinger equation 
\be
\label{SE}
-{d^2 \Psi(y) \over dy^2} + \Lambda^2(e^{-2(y- y_{eq})} - 2e^{-(y - y_{eq})})\Psi(y) = \epsilon \Psi(y)
\ee
where
\be
\label{lambda}
 \Lambda^2= {2 \mu D_e\over \beta^2\hbar^2}
\ee
and 
\be
 \epsilon= {2 \mu E\over \beta^2\hbar^2}
\ee
where $\mu$ is the reduced mass of the system. The Hamiltonian operator in equation (\ref{SE}) is given by
\be
\label{Morse-Hamiltonian}
\hat H_M = -{d^2  \over dy^2} + \Lambda^2(e^{-2(y- y_{eq})} - 2e^{-(y - y_{eq})})
\ee
Since the Schr$\ddot {o}$dinger equation  is exactly solvable, the process of factorization can be completely made and the whole hierarchy can be  evaluated. The
result is, \cite{Drigo2}
\br
V_{n+1}(y) = \Lambda^2( e^{-2(y - y_{eq})} - 2e^{-(y - y_{eq})}) + 2n \Lambda e^{-(y - y_{eq})}
\nonumber
\er
\br
w_{n+1}(y) = - \Lambda e^{-(y - y_{eq})} + (\lambda - {2n + 1\o 2})  \nonumber
\er
\br
\label{non-confined energy}
\epsilon_0^{(n+1)} = -(\Lambda - {2n + 1\o 2})^2. 
\er
\\
In addition we can calculate exactly the wave functions. The ground state wave function is evaluated by using equation (\ref{eigenfunction}) and is given by 
\be
\label{Psi1}
\Psi_0^{(1)}(y) \propto exp\left( {-\Lambda e^{-(y - y_{eq})}\;  -y(\Lambda -\h)} \right).
\ee
This wave function is then used to evaluate, through the SQM algebra,  the  wave function for the excited states by using equation (9); the first excited state is  given by
\be
\label{Psi2}
\Psi_1^{(1)}(y) \propto -2 \left( {\Lambda e^{-(y - y_{eq})}-\Lambda +1} \right) exp\left( {-\Lambda e^{-(y - y_{eq})}\;  -y(\Lambda - {3\o 2})} \right)
\ee
and the second excited state is given by
\br
\label{Psi3}
\Psi_2^{(1)}(y) \propto & \left( {  4\Lambda^2 e^{-2(y - y_{eq})} -8\Lambda^2 e^{-2(y - y_{eq})} +  12\Lambda e^{-(y - y_{eq})}-\Lambda +4\Lambda^2 - 14\Lambda +12
} \right). \\
 & exp\left( {-\Lambda e^{-(y - y_{eq})}\;  -y(\Lambda - {5\o 2})} \right) \;.\nonumber
\er
The wave functions given in (\ref{Psi1}), (\ref{Psi2}) and (\ref{Psi3}) are obtained to the exact system, non-confined or with no hydrogen bond. We refer to them as $ {\Psi_n}_{(non-confined)}$. For the confined system a potential barrier is introduced meaning that the wave functions must be zero outside the barrier.  As the system is no longer exact, it  is necessary to use an approximative method, which we choose to be the variational method.  Thus, based on previous results, \cite{Drigo1} and \cite{Silva}, the eigenfunctions are multiplied by the term  $(y_{max} - y)(y-y_{min})$, i.e., the wave functions become zero at the edges of the confinement region. For the ground state, this approach is equivalent to introduce two additional terms in the superpotential $w_1$, which corresponds to an infinite barrier in the confinement border, $V_1(y_{max})=V_1(y_{min}) = \infty$. 

Adopting this procedure, the wave functions are defined for the region of space in which we are interested and, consequently, the probability of finding the hydrogen outside the confinement region must be zero. Thus, the trial wave functions to be used in the variational method are given by
 \be
 \label{Psi-confined}
{\Psi_n^{(1)}}_{(confined)} \propto {\Psi_n^{(1)}}_{(non-confined)} \; .(y_{max} - y)(y-y_{min})
\ee
and we will only need the cases of $n=0,1,2$. 

The energies for the confined case are obtained by solving
\be
\label{energylevels-confined}
E_n = {\int_{y_{min}}^{y_{max}}{\Psi_n^* \hat H_M \Psi_n dy}\over  {\int_{y_{min}}^{y_{max}}{\Psi_n^*\Psi_n dy}}}
\ee
where $\hat H_M$ is the Hamiltonian of the system given by equation (\ref{Morse-Hamiltonian}) and $\Psi_n$  are the trial wave functions given by equation (\ref{Psi-confined}), where  $ {\Psi_n}_{(non-confined)}$ are given by (\ref{Psi1}), (\ref{Psi2}) and (\ref{Psi3}). The integration limits are the covalent radius of the atom for which the hydrogen is chemically bound ($x_{min}$) and the van der Waals radius complementary of hydrogen bond ($x_{max}$). The results of energy are presented in $cm^{-1}$ units (inverse wavelength) to facilitate comparison with the spectroscopic results.
\section*{\sffamily \large RESULTS }
The hydrogen bonds studied involve the NH and OH groups. Tables 1 and 2 below list the necessary parameters, obtained experimentally,  \cite{Janoschek}-\cite{Bondi}.
\vskip .3cm
\noindent {\bf Table 1.} Parameters used in the NH and OH groups energy evaluations, \cite{Janoschek}.
\vskip .1cm
\begin{center}
\noindent \begin{tabular}{lcccc}  \hline 
\multicolumn{1}{l} { } & 
\multicolumn{1}{l} { } & 
\multicolumn{1}{c} {$NH$} &
\multicolumn{1}{c} { } & 
\multicolumn{1}{c} {$OH$} \\ \hline \\
Equilibrium distance ($x_{eq}$)&& 1.0362 \AA&& 0.9697 \AA\\  
Harmonic vibrational wavenumber ($\omega$)&& $3282 cm^{-1}$&& $3738 cm^{-1}$\\ 
Energy dissociation ($D_e$) & &3.47 eV& &4.39 eV\\  
Reduced mass ($\mu$)&&$1.5614.10^{-27} kg$	&&$1.5746 .10^{-27} kg$\\ 
&& \\ \hline 
\end{tabular}\\
\end{center}
\vskip .3cm
\noindent {\bf Table 2.} Covalent and van der Waals radius used for the nitrogen and oxygen atoms, \cite{Cordero}, \cite{Bondi}.
\begin{center}
\noindent \begin{tabular}{lcc} \hline
\multicolumn{1}{l} { } & 
\multicolumn{1}{c} {$N$} &
\multicolumn{1}{c} {$O$} \\ \hline \\
Covalent radius ($x_c$)& 0.71 \AA& 0.66 \AA\\  
van der Waals radius ($x_{vw}$) & 1.55\AA &1.52 \AA\\  
&& \\ \hline 
\end{tabular}\\
\end{center}
\vskip .3cm
\subsection*{\sffamily \large Results for NH group}
\subsubsection*{\sffamily \normalsize Non-confined case}
The $\beta$ parameter in Morse potential is determined using the values given in Table 1 through the relation (\ref{beta}) and for NH group it is given by  $2.3168 .10^{10} m^{-1}$. The value of the constant $\Lambda$ for NH group, given by equation (\ref{lambda}), is $17.0534$.

In the situation free of any confinement, the calculation of the transition from the first excited state to the ground state is made according to equation (\ref{non-confined energy}). In this case, in spectroscopic units, this value corresponds to $\Delta E_{1\rightarrow 0}=E_1-E_0 = 3089 cm^{-1}$ and it should be checked by means of spectroscopy. By IR spectroscopy with polypeptides, an absorption band in $3090 cm^{-1}$ is observed and it has been described as a vibration frequency of the NH  molecular group, \cite{Badger}. This value is very close to the calculated and can be identified, according to the proposed model, as vibration of NH without formation of hydrogen bonding.
\subsubsection*{\sffamily \normalsize Confined case} 
When the NH group form molecular hydrogen bonding, the vibration frequency is shifted. An interesting system where  this effect can be observed is in proteins. These may be in the form of $\alpha$-helix, $\beta$-sheet or random.  In this helicoidal format the hydrogen bonding is between the amine group of an amino acid and the oxygen of the carboxyl group of another amino acid. Thus, proteins consist in a good system to verify the quantum confinement effect in hydrogen bonding. It is found in the literature that proteins  in $\alpha$-helix  format present hydrogen bond lengths around $2.80 \AA$, \cite{Voet}. Considering the uncertainty of experimental results, the values of $2.75 \AA$, $2.80 \AA$ and $2.85 \AA$ were used as the maximum distance of the hydrogen oscillation,  according to the equation (\ref{xmax}). Solving equation (\ref{energylevels-confined}) numerically, the results for the emission/absorption for proteins in $\alpha$-helix are indicated below in Table 3.\\

\noindent {\bf Table 3.} Energy levels of proteins with hydrogen bonding, for different values of $x_{max}$ in $(cm^{-1})$ units. 

\begin{center}
\noindent \begin{tabular}{lccccc} \hline
\multicolumn{1}{c} {$x_{max}$ } & 
\multicolumn{1}{c} {$E_0$} &
\multicolumn{1}{c} { } & 
\multicolumn{1}{c} {$E_1$} & 
\multicolumn{1}{c} { } & 
\multicolumn{1}{c} {$\Delta E_{1\rightarrow 0}$} \\ \hline
$2.75 \AA$ &-26105 && -22643 && 3461 \\ \hline
$2.80 \AA$ &-26245 && -22835 & &3410 \\ \hline
$2.85 \AA$ &-26306&	&-23010&	&3296\\  \hline
\end{tabular}\\
\end{center}

The hemoglobin, a metalloprotein present in the red blood cells of all vertebrates, is constituted of four $\alpha$-helix chains. IR spectroscopy realized to verify the alterations caused by $\beta$-thalassemia disorder, a reduction or absence synthesis of the beta chains of hemoglobin, \cite{Liu}, showed peaks around $3050 cm^{-1}$  and $3290 cm^{-1}$, that can be identified as vibrations of the NH group in the situations free of confinement and confined, or with or without hydrogen bond, respectively. By considering $2.85 \AA$ as the maximum distance of the hydrogen oscillation,  the comparison of our theoretical result for the confined case ($3296 cm^{-1}$) with these results shows an error of $0.18\%$. Thus we expect that other proteins in $\alpha$-helix format present absorption peaks around $3296 cm^{-1}$. In fact, other proteins in the $\alpha$-helix format present transitions verified by IR spectroscopy  that corroborate with our results, as shown below. 

Keratin, for instance, is a protein found in abundance in life,  in the human epidermis, hair, nails, feathers, porcupine quills, etc. IR spectroscopy of virgin human hair (i.e., no chemical intervention) and horse hair, using lithium bromide and hydrochloric acid as reagent, \cite{Baddiel}, showed the keratin absorption bands of hair with and without cuticle. Similar studies were carried out with a dried sample of goose feather followed by a soaked sample using heavy water, \cite{Parker}. The IR spectroscopy studies with some peptides, polyamides and fibrous proteins used swan feather wing and horse hair as a source of keratin, \cite{Bradbury}. 
 
Other protein having $\alpha$-helix  structure is rhodopsin, which can be found in the eye retina. IR spectroscopy with bovine rhodopsin  Langmuir-Blodgett films, \cite{Pepe}, showed an absorption peak in $3295 cm^{-1}$; the theoretical result presents $0.03\%$ error relative to this experimental result.

The sodium-potassium pump (Na+/K+-ATPase) present in cells is another protein which contains several $\alpha$-helices, \cite{Fringeli}. The IR spectroscopy carried out in an aqueous medium of three different ionic compositions presented absorption peaks; in all types of solvent there is a peak located in $3285 cm^{-1}$; the theoretical result presents $0.33\%$ error relative to this experimental result.

Table 4 summarizes the results discussed above.
\vskip .3cm
\noindent {\bf Table 4.} Results for absorption/emission values $\Delta E_{1\rightarrow 0}= E_1- E_0$, in $(cm^{-1})$ units for the non-confined and confined cases with deviations relative to the experimental result.
\begin{center}
\noindent \begin{tabular}{lccccccccccc} \hline \\
\multicolumn{1}{c} { } & 
\multicolumn{1}{c} { } &
\multicolumn{1}{c} {Non-Confined}&
\multicolumn{1}{c} { } & 
\multicolumn{1}{c} {Confined}& 
\multicolumn{1}{c} {Deviation from  }  \\
&&(no hydrogen& &(with hydrogen  & experimental  \\ 
&&bond)& &bond)  & (\%)  \\\hline
Theoretical &&3089&& 3296&& \\
Hemoglobin with $\beta$-thalassemia disorder, \cite{Liu} &&	3050 &&	3290& 0.18	\\
Normal Hair, \cite{Baddiel}	&&	3066 &&	3300&0.12	\\
Hair without cuticle, \cite{Baddiel}	&&3062	&&3295& 0.03	\\
Horse Hair,	\cite{Baddiel}&&-	&&3293&0.09\\
Goose feather, \cite{Parker}	&&3100	&&3300&	0.12\\
Swan Feather Wing, \cite{Bradbury}	&&3070 &&	3290&0.18 \\
Horse Hair, 	 \cite{Bradbury}&&3080	&&3295&0.03	\\
Rhodopsin (Retina eye), \cite{Pepe}&&-&&	3295&0.03	\\ 
Sodium-Potassium Pump, \cite{Fringeli}&&-&&	3285&0.33	\\ \hline

\end{tabular}
\end{center}

\subsection*{\sffamily \large Results for OH group}
\subsubsection*{\sffamily \normalsize Non-confined case} 

In the case of OH group, the transition from the second to the first excited state ($\Delta E_{2\rightarrow 1}= E_2-E_1$) is also studied. According to the formalism described, the $\beta$ parameter, equation (\ref{beta}), for the OH group is $2.3558.10^{10} m^{-1}$ and the $\Lambda$ constant value is 18.9424. In the situation free of any type of confinement, i.e. with no hydrogen bond, the transitions between adjacent states were evaluated by using equation (\ref{non-confined energy}); the results  are shown in Table 5.\\

\noindent {\bf Table 5.} Eigenvalues of energy to OH obtained by the model described in $cm^{-1}$ units.

\begin{center}
\noindent \begin{tabular}{lccccccccccccc} \hline
\multicolumn{1}{c} { } & 
\multicolumn{1}{c} {$E_0$} &
\multicolumn{1}{c} { } & 
\multicolumn{1}{c} {$E_1$} & 
\multicolumn{1}{c} { } & 
\multicolumn{1}{c} {$E_2$} &
\multicolumn{1}{c} { } & 
\multicolumn{1}{c} {$\Delta E_{1\rightarrow 0}$}&
\multicolumn{1}{c} { } & 
\multicolumn{1}{c} {$\Delta E_{2\rightarrow 1}$}&
\multicolumn{1}{c} { } &  \\ \hline
&-33560&&	-30019	&&-26676&&	3541	&&3343\\  \hline

\end{tabular}\\
\end{center}
\subsubsection*{\sffamily \normalsize Confined case}
Carboxylic acid dimers are held by hydrogen bonds between the hydroxyl.  By means of neutron diffraction, differently to the X-ray diffraction, one can obtain  the distances between the hydrogen atom and the other atoms that compose the network, \cite{Jeffrey}. Figure 2  shows the distance between the hydrogen, linked to the donor atom, and the acceptor atom of the bond. Thus, to find the maximum distance of oscillation is necessary to sum the equilibrium distance of the OH bond to the length of hydrogen bond, i. e., the maximum distance is given by
\be
\label{xmax-2}
x_{max} = x_{eq} + l_{H-O} - x_{vw}\;.
\ee
 	
Three types of carboxylic acids in the form of dimers, are studied in this work, the lengths of the hydrogen bonds are reported in Table 6.

\vskip 1cm

\noindent {\bf Table 6.} Length of hydrogen bonds in dimers of different acids, \cite{Jeffrey}.

\noindent \begin{tabular}{lccccccccc} \hline
\multicolumn{1}{c} { } & 
\multicolumn{1}{c} { } &  
\multicolumn{1}{c} {Trifluoroacetic acid} &
\multicolumn{1}{c} { } & 
\multicolumn{1}{c} { } & 
\multicolumn{1}{c} {Acetic acid} &
\multicolumn{1}{c} { } & 
\multicolumn{1}{c} { } & 
\multicolumn{1}{c} {Formic acid } & 
\multicolumn{1}{c} { } \\ \hline 
Bond length ( $l_{H-O}$)	&&1.658 \AA&&&	1.642\AA&&& 1.660\AA \\  \hline
\end{tabular}\\
\vskip 1cm
Using the values of Table 6 to find the maximum distance of oscillation of hydrogen according to the equation (\ref{xmax-2}) and solving equation (\ref{energylevels-confined}) with wave functions given by (\ref{Psi-confined}) and (\ref{Psi1})-(\ref{Psi3}) we find the value of emission/absorption energies for carboxylic acid dimers, shown in Table 7.
\vskip 1cm
\noindent {\bf Table 7.} Energy levels of different systems, in  $(cm^{-1})$ units.

\noindent \begin{tabular}{lcccccccccccccc} \hline
\multicolumn{1}{c} { } & 
\multicolumn{1}{c} { } & 
\multicolumn{1}{c} {$E_0$}&
\multicolumn{1}{c} { } & 
\multicolumn{1}{c} { } & 
\multicolumn{1}{c} {$E_1$}& 
\multicolumn{1}{c} { } & 
\multicolumn{1}{c} { } & 
\multicolumn{1}{c} {$E_2$} & 
\multicolumn{1}{c} { } & 
\multicolumn{1}{c} { } & 
\multicolumn{1}{c} {$\Delta E_{1\rightarrow 0}$}&
\multicolumn{1}{c} { } & 
\multicolumn{1}{c} { } & 
\multicolumn{1}{c} {$\Delta E_{2\rightarrow 1}$} \\ \hline
Trifluoroacetic acid	&&	-32953 &&&		-29282 &&&	-26270 &&&	3671	&&&2944\\
Acetic acid	&&-32767	&&&-29342	&&&-26270	&&&3425 &&&	3072\\
Formic acid	&&-32973&&&	-29277	&&&-26342	&&&3696 &&&	2935\\  \hline

\end{tabular}\\

The trifluoroacetic acid was analyzed in neon and argon matrices, in its monomeric form, in cyclic dimers and also in the deuterated form ($CF_3COOD$) by means of IR spectroscopy, \cite{Redington}. In neon matrix the dimer of trifluoroacetic acid presented emission/absorption of $2922 cm^{-1}$, described as the vibration of the OH molecular group, because when the acid is deuterated this peak disappears. The theoretical value presents $0.75\%$ of error relative to this experimental value, considering the transition between the second excited state and the first excited state, $\Delta E_{2\rightarrow 1}= E_2-E_1$. 

Acetic acid was also analyzed by IR spectroscopy, \cite{Bratoz}. There were treated several dimers of carboxylic acids. The values of $3067 cm^{-1}$ in gas phase, $3100 cm^{-1}$ carbon tetrachloride solution ($CCl_4$) and $3080 cm^{-1}$ in liquid phase  agree with a transition $\Delta E_{2\rightarrow1}$ (Table 7) of the confined case, with theoretical errors smaller than $0.9\%$, obtained for the case where the acid was in  $CCl_4$ solution.

IR spectroscopy was also used to analyse the formic acid photocatalytic decomposition process, \cite{Arana}. When it is analyzed on the catalyst surface of $TiO_2$, without any additive, an absorption peak near $3698 cm^{-1}$ can be noted, compatible to the transition $\Delta E_{1\rightarrow 0}$ (Table 7).  This peak is attributed by the authors to the OH group, which during the reaction is displaced to values below $3668 cm^{-1}$. This can be explained by the dissociation of the OH group during decomposition, preventing the formation of hydrogen bond and weakening the bonds still existing. 

The IF spectroscopy of photocatalytic decomposition of formic acid was analyzed in \cite{Chen}, showing transitions identified as the vibration of the OH group, which agrees with the transition of $\Delta E_{1\rightarrow 0}$ (Table 7).  
Table 8 summarizes the results discussed above.
\vskip 1cm
\noindent {\bf Table 8.} Results for absorption/emission values $\Delta E_{1\rightarrow 0}= E_1- E_0$, in $(cm^{-1})$ units for the non-confined and confined cases with deviations relative to the experimental result.
\begin{center}
\noindent \begin{tabular}{lcccccccccccccc} \hline
\multicolumn{1}{c} { } & 
\multicolumn{1}{c} { } & 
\multicolumn{1}{c} { } & 
\multicolumn{1}{c} {$\Delta E_{1\rightarrow 0}$}&
\multicolumn{1}{c} { } & 
\multicolumn{1}{c} { } & 
\multicolumn{1}{c} {$\Delta E_{2\rightarrow 1}$} & 
\multicolumn{1}{c} { } & 
\multicolumn{1}{c} { } & 
\multicolumn{1}{c} {Deviation from  }  \\
&&&&&& & &  & experimental (\%)  \\  \hline
Trifluoroacetic acid, \cite{Redington}	&&&	-	&&&2922&&&0.75\\
Theoretical&&&	3671	&&&2944\\ \hline
Acetic acid, \cite{Bratoz}	&&&  &&&	 \\
$\;\;\;\;\;$Gas phase &&& - &&&	3067&&&0.16\\
$\;\;\;\;\;$In $CCl_4$ solution&&& - &&&	3100&&&0.90 \\
$\;\;\;\;\;$Liquid phase&&& - &&&	3080&&&0.26 \\
Theoretical&&&3425 &&&	3072\\ \hline
Formic acid,  \cite{Arana}	 &&&3698 &&&	- &&&0.05\\ 
Formic acid,  \cite{Chen}	 &&&3697 &&&	- &&&0.03\\ 
Theoretical&&&3696 &&&	2935\\  \hline
\end{tabular}\\
\end{center}
\section*{\sffamily \large CONCLUSIONS}
In this work the quantum confinement effect in hydrogen bonds explains the shift in the emission spectrum of NH and OH groups. 
In all the cases mentioned the distortion in the spectrum when the bond is formed is considerable and measurable, and is easily distinguished from the non-confined case, with no hydrogen bond. This shift occurs because the hydrogen bond imposes a barrier to the hydrogen and prevents it from having freedom in its oscillatory motion. 

Quantitative results obtained for the vibrational transitions of the groups mentioned agree with the experimental results reported in the literature, reenforcing the applicability of the proposal and the reliability of this quantum mechanical methodology. The errors between the theoretical values entered by the variational method and the experimental data analyzed are small, the largest percentage error obtained in the calculations is smaller than $1.0\%$.  

The test wave functions used in the confined case, obtained with the aid of the SQM formalism, are appropriate to describe the phenomenon. As a final result of the construction of these wave functions, the polynomial terms  are multiplied by the wave functions of the free system, with no confinement. This term of confinement is sufficient to obtain the numerical results presented.

As a final remark, the treatment proposed here can be extended to the study of other molecular groups, as FH and CH, involved in hydrogen bonds.

\clearpage
\begin{figure}
\centering
\includegraphics[width=0.8\textwidth]{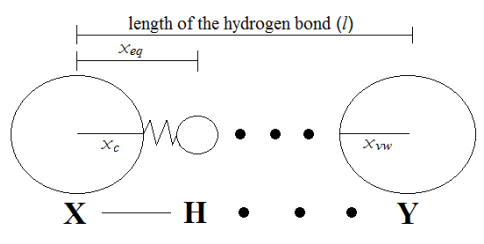}
\caption{\label{label} Schematic one-dimensional model, where $x_c$ is the covalent radius of the donor atom X,  $x_{vw}$ is the van der Waals radius of the  acceptor atom Y and $x_{eq}$ is the equilibrium distance of covalent bond. }
\end{figure}

\begin{figure}
\centering
\includegraphics[width=0.6\textwidth]{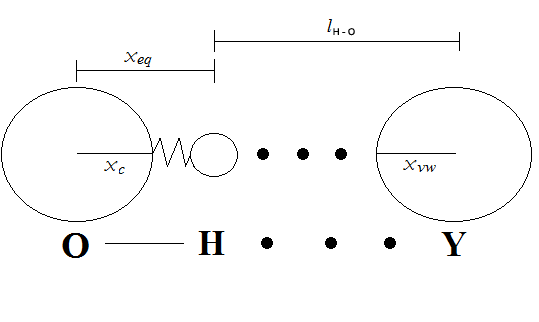}
\caption{\label{label} Schematic one-dimensional model for hydrogen bond for the group OH, remarking the length of hydrogen bond, \cite{Jeffrey}. }
\end{figure}

\end{document}